\documentclass[pra,aps,showpacs,twocolumn,floatfix]{revtex4}
\usepackage{epsf}
\usepackage[dvips]{color}
\newcommand {\be}{\begin{equation}}
\newcommand {\ee}{\end{equation}}
\newcommand {\bea}{\begin{eqnarray}}
\newcommand {\eea}{\end{eqnarray}}

\begin{document}

\title{Endpoint thermodynamics of an atomic Fermi gas
subject to a Feshbach resonance}
\author{L.~D. Carr, R. Chiaramonte, and M. J. Holland\\}
\affiliation{JILA, National Institute of Standards and Technology
and University of Colorado, Boulder, CO 80309-0440}
\date{\today}

\begin{abstract}
The entropy and kinetic, potential, and interaction energies of an
atomic Fermi gas in a trap are studied under the assumption of
thermal equilibrium for finite temperature. A Feshbach resonance
can cause the fermions to pair into diatomic molecules. The
entropy and energies of mixtures of such molecules with unpaired
atoms are calculated, in relation to recent experiments on
molecular Bose-Einstein condensates produced in this manner. It is
shown that, starting with a Fermi gas of temperature $T= 0.1
T_F^0$, where $T_F^0$ is the non-interacting Fermi temperature, an
extremely cold degenerate Fermi gas of temperature $T \lesssim
0.01 T_F^0$ may be produced without further evaporative cooling.
This requires adiabatic passage of the resonance, subsequent
sudden removal of unpaired atoms, and adiabatic return.  We also
calculate the ratio of the interaction energy to the kinetic
energy, a straightforward experimental signal which may be used to
determine the temperature of the atoms and indicate condensation
of the molecules.
\end{abstract}


\maketitle

\section{Introduction}
\label{sec:intro}

The formation of cold diatomic molecules from an atomic Fermi gas
offers one the opportunity to study the Bose-Einstein condensate
to Bardeen-Cooper Schrieffer (BEC to BCS) crossover in an atomic
system~\cite{randeria1995,holland2001,timmermans2001,ohashi2002}.
Cold atomic gases present several advantages in investigations of
this long-standing problem: they are impurity-free; their
temperature, density, and interaction strength can be tuned over
wide ranges; and they are easily manipulated with lasers.  The key
to obtaining superfluid states such as BEC and BCS is to cool the
gas to a sufficiently low temperature~\cite{density}.

In recent experiments, Feshbach resonances have been used to
transform a cold or degenerate Fermi gas into a molecular
BEC~\cite{greiner2003,zwierlein2003,jochim2003b,regal2004,kinast2004,bourdel2004}.
A Feshbach resonance is a resonance between an unbound atomic
state and a bound molecular state brought about by the Zeeman
effect, in practice by changing a uniform magnetic field over the
region of the gas~\cite{vogels1}. The molecular pairing mechanism
is a few-body effect brought about by the Feshbach resonance. The
BCS pairing mechanism is a many-body effect due to an instability
at the Fermi surface.  This interplay between few and many body
physics is an outstanding problem.  Various theories have been
proposed to model the crossover region and molecule
formation~\cite{holland1,koehler2003b,ohashi2003,petrov2003b,kokkelmans2004,holland2004,diener2004,falco2004,drummond2004,andreev2004}.
However, in order to study the crossover thoroughly it is
desirable to have clearly defined and demonstrated endpoints.

In the following, basic thermodynamic quantities related to the
endpoints of this problem are calculated:  in particular, the
kinetic energy, potential energy, interaction energy, and the
entropy.  Two cases are considered: (a) a degenerate atomic Fermi
gas of two spin states interacting via a weakly attractive contact
potential; and, (b) an atom-molecule mixture in which the atoms
and molecules interact both with themselves and with each other
via a repulsive contact potential, as was the case in the
experiment of Ref.~\cite{greiner2003}. The atomic interaction
parameter is given by $k_F^0 |a|$, where $k_F^0$ is the
non-interacting Fermi momentum and $a$ is the $s$-wave scattering
length. The $s$-wave scattering length is a function of the
magnetic field in the region of the resonance and typically
changes over many orders of magnitude as well as in sign. Case (a)
treats the parameter regime $a<0$ and $k_F^0 |a|\ll 1$, for which
a weakly interacting BCS-type transition is possible. Case (b)
treats the regime $a>0$ and $k_F^0 |a|\ll 1$, in which case the
molecules may Bose condense. In case (b) both a Boltzmann-like
molecular gas of $T/T_{\mathrm{BEC}}\gg 1$ and a degenerate
molecular gas of of $T/T_{\mathrm{BEC}}\ll 1$ are considered. Two
applications which demonstrate the usefulness of these
calculations in relation to present experiments are illustrated.

Firstly, we present a straightforward signal for the onset of
condensation in the molecules. The ratio of the interaction energy
to the kinetic energy is an established
observable~\cite{bourdel2003}.  The power law in this ratio
changes as the molecules are condensed.  Moreover, this quantity
can be used to determine the temperature, which, particularly in
case (b) for $T\ll T_{\mathrm{BEC}}$, would be difficult to
determine for the atoms from momentum distributions alone, since
they are pushed to the edges of the molecular cloud by the
molecular mean field, as we shall show.  This ratio is thought to
be a universal constant in the unitarity limit, {\it i.e.}, when
$k_F^0|a|\gtrsim 1$, where it is called $\beta$ in the
literature~\cite{ohara1,ho2004}.

Secondly, we propose a new way to obtain a sufficiently low
temperature for a weakly interacting BCS transition.  A recent
proposal~\cite{carr36} showed that adiabatic transition of the
resonance from degenerate molecules to atoms leads to cooling.
This follows from the fact that the entropy of an ideal fermionic
gas $S_{\mathrm{a}}\propto T$ while the entropy of an ideal
bosonic gas $S_{\mathrm{m}}\propto T^3$~\cite{huang1}. Holding the
entropy constant results in a decreased temperature for
sufficiently small $T$.  The addition of weak interactions does
not change this fact~\cite{carr36}. In the experiment of
Ref.~\cite{greiner2003}, it was found that, after tuning the atoms
over the resonance so that they formed molecules, about 15\% of
the atoms remained unpaired~\cite{narrow}. In such an
atom-molecule mixture, at low temperature, the majority of the
entropy resides in the atoms. Therefore, suddenly removing the
atoms is the same as removing entropy.  The technique of removing
atoms while leaving the molecules in place has been experimentally
demonstrated~\cite{xu2003}.  Adiabatically dissociating the
molecules then leads to an extremely cold fermionic gas of
temperatures $T\lesssim 0.01 T_F$, where $T_F$ is the Fermi
temperature. We term this cooling method \emph{entropic cooling}.
Given the experimental limitations of evaporative cooling, which
for a BEC is $T \sim 0.25 \,T_{\mathrm{BEC}}$ and for a Fermi gas
is $T\sim 0.05 \,T_F$ (see, for example, Refs.~\cite{chevy2002}
and~\cite{hadzibabic2003,greiner2003}, respectively), entropic
cooling provides an attractive alternative.

We note that a recent experiment has successfully traversed the
crossover from a molecular BEC to an atomic Fermi gas in a
sufficiently adiabatic manner to achieve entropic
cooling~\cite{bartenstein2004}.

The presentation may be outlined as follows.  In
Sec.~\ref{sec:model} the model is presented. In
Sec.~\ref{sec:negative} the thermodynamic quantities are
calculated for a degenerate Fermi gas, in some detail.  In
Sec.~\ref{sec:positive} the same quantitites are presented for the
atom-molecule mixture at temperatures both above and below
$T_{\mathrm{BEC}}$. In Sec.~\ref{sec:application} the two
experimental applications are considered: a signal for the onset
of molecular BEC; and the use of entropic cooling to achieve
temperatures sufficiently low for a BCS transition. Finally, in
Sec.~\ref{sec:conclusion} we conclude.

\section{The Model}
\label{sec:model}

Our model consists of the following assumptions.  Firstly, it is
supposed that for negative scattering length $T>T_{\mathrm{BCS}}$
and $k_F^0 |a| \ll 1$. This assures that the system is far from
resonance and that there is no superfluid phase.  Secondly, for
positive scattering length the dilute limit $\sqrt{\bar{n}_m
a_{\mathrm{mm}}^3} \ll 1$ is assumed for the molecules, where
$\bar{n}_m$ is the average molecular density and the
molecule--molecule and atom--molecule scattering lengths
$a_{\mathrm{mm}},\, a_{\mathrm{am}} \propto a$, with $a$ the
s-wave atom--atom scattering length.  The proportionality
constants are taken to be 0.6 and 1.2, respectively, in accord
with Petrov {\it et al.}~\cite{petrov2003b}, where it is
calculated explicitly in certain limits from the four body
problem. Thirdly, it is assumed that the atoms and molecules can
be treated as distinct entities: the former are fermions and the
latter are bosons, whose thermodynamic properties can be
calculated separately. Fourthly, it is assumed that the system is
in thermal equilibrium.  However, it is not necessary to require
chemical equilibrium.  The former requires a sufficiently high
rate of elastic two body collisions ($A+A\leftrightarrow A+A$,
$M+A\leftrightarrow M+A$); as the latter is brought about by
three-body collisions ($A+A+A\leftrightarrow M+A$), it may be
substantially slower. This indeed appears to be case in certain
experiments~\cite{greiner2003,kokkelmans2004,narrow,depalo2004}.

Fifthly, it is assumed that the atom--atom coupling for binary
interactions is given by the $s$-wave limit $g=2\pi\hbar^2 a/m_r$,
where $m_r=m_a/2$ is the reduced mass.  The atom--molecule and
molecule--molecule coupling is then $g_{\mathrm{am}}=0.9 g$ and
$g_{\mathrm{mm}}=0.3 g$ according to Ref.~\cite{petrov2003b}.
Sixthly, it is assumed that both atoms and molecules are subject
to the same isotropic harmonic trapping frequency $\omega$.  Thus
their densities shall also be isotropic, as well as all other
thermodynamic quantities~\cite{anisotropic}.  Seventhly, it is
assumed that for negative scattering length there are an equal
number of spin up and spin down atoms.  It is now possible to
change the scattering length from negative to positive,
selectively remove excess atoms of both spin states, and then
change the scattering length back to negative, thereby resulting
in the precise balance assumed here~\cite{xu2003}.  This results
in a pleasant symmetry in the thermodynamic expressions, as well
as avoiding Fulde-Ferrell-Larkin-Ovchinnikov phases when a BCS
transition is sought (see Ref.~\cite{combescot2001} and references
therein). Eighthly, the local density and semi-classical
approximations, which are valid for $k_B T \gg 1/\rho(E_F)$, are
assumed to hold, where $\rho(E_F)$ is the density of states at the
Fermi surface.

The critical temperature for condensation of an ideal gas shall be
denoted as \be k_B T^0_{\mathrm{BEC}}\equiv \hbar \omega
\left[\frac{N_m}{\zeta(3)}\right]^{1/3}\, ,\ee where $N_m$ is the
number of molecules and $\zeta$ is the Riemann Zeta function. The
shift in this temperature caused by interactions may be calculated
to be~\cite{dalfovo1} \be \frac{\delta T}{T_{\mathrm{BEC}}^0} =
-1.3 \frac{a_{\mathrm{mm}}}{l_{ho}}N_m^{1/6}\ee in a harmonic trap
for the assumed regime of $\sqrt{\bar{n}_m a_{\mathrm{mm}}^3}\ll
1$, where \be l_{ho}\equiv\sqrt{\hbar/(m_m\omega)} \ee is the
harmonic oscillator length.  Therefore a ninth assumption is that
this shift is small, so that it may be neglected in our
calculations~\cite{metikas}. This is consistent with the second
assumption above. Typical shifts in experiments are less than
10\%. Similarly, the Fermi temperature will be denoted by its
ideal value of \be k_B T^0_F\equiv \hbar \omega (3 N_a)^{1/3}\,
,\ee where $N_a$ is the number of atoms.

Lastly, for $T\ll T_{\mathrm{BEC}}$, the Thomas-Fermi profile
shall be assumed for the bosonic molecules: \be
n^{TF}_m(r)=\frac{\mu_m}{g_{\mathrm{mm}}}\left(1-\frac{r^2}{R_m^2}\right),\,\,\,\,\,r\leq
R_m \,,\ee \be \mu_m=\frac{1}{2} m_m \omega^2 R_m^2\, ,\ee where
the subscript $m$ refers to molecules, $m_m=2m_a$ and $\mu_m$ is
the molecular chemical potential.  Note that $n^{TF}_m(r)=0$ for
$r \geq R_m$.

\section{Negative scattering length: atoms only}
\label{sec:negative}

We first consider the case for which the scattering length is
negative, so that there are no molecules.  We calculate the
entropy and the energy. The energy is divided into kinetic,
potential, and interaction parts. For an isotropic harmonically
trapped cold atomic gas, the potential energy \be
V(r)=\frac{1}{2}m_a\omega^2 r^2 \ee represents the trap energy and
\be U_{\mathrm{mf}}(r) = \frac{1}{2}g n_a(r) \ee is the
interaction or mean field energy, where $n_a(r)$ is the total mean
field particle density, {\it i.e.}, including both spin states,
and $g$ is the interaction strength. The expressions are derived
in some detail in this, the simplest case, in order to serve the
reader as a model for similar calculations which are only sketched
in the following sections.

\subsection{Energy}
\label{ssec:atomenergy} The expressions for the energy in the
local density approximation are given by integrals over phase
space of the form \bea E_{\mathrm{kin}}&=&2\int \frac{d^3r
d^3p}{(2\pi\hbar)^3}\,\frac{p^2}{2m_a}\,\nu(r,p)
\, , \label{eqn:KE}\\
E_{\mathrm{pot}}&=&2\int \frac{d^3r
d^3p}{(2\pi\hbar)^3}\,\frac{1}{2}m_a\omega^2 r^2\,\nu(r,p)
\, , \label{eqn:PE}\\
E_{\mathrm{int}}&=&\int d^3r\,g \left[\frac{1}{2}n_a(r)\right]^2
\, , \label{eqn:INT}\eea where \be
\nu(r,p)\equiv\{\exp[\beta(E(r,p)-\mu)]+ 1\}^{-1}
\label{eqn:nu}\ee is the mean occupation number, \be
\beta\equiv\frac{1}{k_B T}\, ,\ee and \be
E(r,p)\equiv\frac{p^2}{2m_a}+V(r)+U_{\mathrm{mf}}(r)~\label{eqn:ephasespace1}\ee
is the total energy in phase space. The factor of 2 in front of
the integrals in Eqs.~(\ref{eqn:KE}) and~(\ref{eqn:INT}) is due to
the number of spin states.

One may obtain self-consistent expansions to first order in $k_F^0
|a|$ and second order in $T/T_F^0$, where \be
k_F^0\equiv\sqrt{2m_a k_B T_F^0}/\hbar\, .\ee is the Fermi
momentum defined with respect to an ideal gas. The key is to set
$n_a(r)=n_a^{(0)}(r)$ in $U_{\mathrm{mf}}(r)$ and
Eq.~(\ref{eqn:INT}), where \bea n_a^{(0)}(r)&\equiv&
\frac{1}{4}\left(\frac{2m_a}{\pi\hbar^2\beta}\right)^{3/2}
g_{3/2}\left(-e^{u}\right)\, ,\label{eqn:density}\\
u&\equiv&\beta\left(\mu_a-\frac{1}{2}m_a\omega^2 r^2\right)\,
,\eea is the non-interacting density profile, which may be
obtained by integrating Eq.~(\ref{eqn:nu}) over momentum space.
The function $g_{n}(x)$ in Eq.~(\ref{eqn:density}) is the Bose
function~\cite{huang1}. The details of the calculation are
presented in App.~\ref{app:fermicorrect}. One finds \bea
E_{\mathrm{kin}}&=&Nk_B T_F^0
\left[\left(\frac{3}{8}+\frac{256}{315\pi^2} k_F^0 |a|\right)
\right.\nonumber\\
&&\left.
+\left(\frac{\pi^2}{4}-\frac{184}{35} k_F^0 |a|\right)
\left(\frac{T}{T_F^0}\right)^2\right]\, ,\label{eqn:KE3}\\
E_{\mathrm{pot}}&=&N k_B T_F^0
\left[\left(\frac{3}{8}-\frac{256}{315\pi^2} k_F^0 |a|\right)
\right.\nonumber\\
&&\left.
+\left(\frac{\pi^2}{4}-\frac{136}{35} k_F^0
|a|\right)\left(\frac{T}{T_F^0}\right)^2\right]\, ,\label{eqn:POT3}\\
E_{\mathrm{int}}&=&-N k_B T_F^0 k_F^0 |a|
\left[\frac{1024}{945\pi^2}
-\frac{32}{35}\left(\frac{T}{T_F^0}\right)^2\right].\,\label{eqn:INT3}\eea

These expressions may be shown to be self-consistent in two ways.
Firstly, one observes that the virial theorem \be
2E_{\mathrm{kin}}-2E_{\mathrm{pot}}+3E_{\mathrm{int}}=0
\label{eqn:virial1}\ee holds up to first order in $k_F^0 |a|$ and
to second order in $T/T_F^0$. Secondly, note that the total energy
at zero temperature for an ideal gas is $E^{\mathrm{total}}_{T=0}
= E_{\mathrm{kin}}+E_{\mathrm{pot}} = \frac{3}{4}N E_F^0$, as may
also be calculated directly from the density of states for an
ideal Fermi gas in a harmonic trap.  The expansions given by
Eqs.~(\ref{eqn:KE3}),~(\ref{eqn:POT3}), and~(\ref{eqn:INT3}) are
accurate to 1\%, 5\%, and 24\% for $k_F^0|a|=0.1$ and
$T/T_F^0=0.1$.  For higher temperature or stronger interaction
strength, the accuracy becomes quite poor, for example 72\%, 35\%,
and 233\% for $k_F^0|a|=0.5$ and $T/T_F^0=0.5$.  In this case it
is preferable to evaluate the energy integrals self-consistently
in the mean field by recursive use of
Eqs.~(\ref{eqn:KE})-(\ref{eqn:INT}) with the density profile \be
n_a(r)=2 \int \frac{d^3p}{(2\pi\hbar)^3}\nu(r,p)\,
,\label{eqn:density2}\ee while holding the total number of atoms,
given by the integration of Eq.~(\ref{eqn:density2}) over volume,
constant.  This gives the exact result, which we have used to
evaluate the accuracy of the expansions.

\begin{figure}[t]
\begin{center}
\epsfxsize=8cm \leavevmode \epsfbox{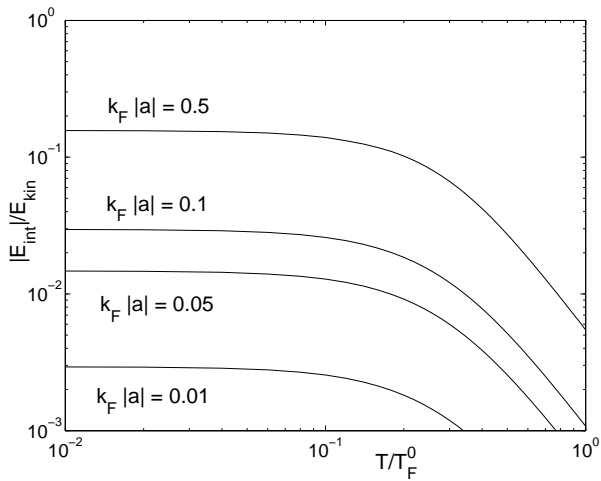}
\caption{\label{fig:1}  Shown is the ratio of the interaction
energy to the kinetic energy as a function of $T/T_F^0$ for a
weakly interacting degenerate Fermi gas.  This ratio is an
established experimental observable~\cite{bourdel2003} that may be
used to measure the degeneracy of the gas.  This ratio, called
$\beta$ in the unitarity limit, {\it i.e.}, when $k_F^0|a|\gtrsim
1$, is thought to be a universal constant in that
regime~\cite{ohara1,ho2004}.  Here, the temperature dependence of
the ratio is illustrated for increasing values of $k_F^0|a|$.}
\end{center}
\end{figure}

In Fig.~\ref{fig:1} is shown the ratio of the interaction energy
to the kinetic energy in the degenerate regime.  This quantity is
a straightforward experimental observable~\cite{bourdel2003} which
can be used to determine the temperature of the system. The figure
shows the self-consistent solution which makes no assumptions on
the smallness of the mean field or the temperature. This is
particularly important for the upper curve, which treats the
almost strongly interacting regime. In a recent work, it was shown
that values of $k_F^0|a|$ likely to lead to a BCS transition are
on the order of one-half~\cite{carr36}.

%

\subsection{Entropy}
\label{ssec:atomentropy}

The general combinatorial expression for entropy is \be S=-k_B
\sum_{\{\lambda\}}p_{\lambda}\ln p_{\lambda}\ee where
$\{\lambda\}$ represents all states of the system and
$p_{\lambda}$ is the probability to be in a given
state~\cite{landau3}. For fermions in the local density
approximation in an isotropic system, \be S=-2 k_B \sum_{\{r,p\}}[
\nu \ln \nu +(1-\nu)\ln(1-\nu)]\, ,~\label{eqn:entropy1}\ee where
the factor of 2 is due to the two atomic spin states and $\nu$ is
defined in Eq.~(\ref{eqn:nu}). Equation~(\ref{eqn:entropy1}) can
be integrated over momentum to obtain \be S(r)=A
\frac{\sqrt{\pi}}{2}\left[ -\frac{5}{2}g_{5/2}\left(-e^u\right)
+u\,g_{3/2}\left(-e^u\right) \right]\, .~\label{eqn:genent}\ee The
details are given in App.~\ref{app:fermicorrect2}. This is the
general expression for the entropy of the fermions.

For an ideal Fermi gas in a harmonic potential, the integral of
Eq.~(\ref{eqn:genent}) can be performed exactly (see
Eq.~(\ref{eqn:myint}) in App.~{\ref{app:fermicorrect}}). One
obtains \bea \frac{S}{k_B} &=& 2\left(\frac{k_B
T}{\hbar\omega}\right)^3\left[\beta
g_3\left(-e^{\beta\mu_a}\right)\right.\nonumber\\
&&\left.-4 g_4\left(-e^{\beta\mu_a}\right)\right] \,
.\label{eqn:exactidealatom}\eea In the degenerate limit,
$\beta\mu_a \gg 1$, the Bose functions may be expanded to obtain
(see Eq.~(\ref{eqn:boseexpand}) in App.~\ref{app:fermicorrect})
\be S = k_B \pi^2 N \left(\frac{T}{T_F^0}\right)^3 (\beta\mu_a)^2
\label{eqn:fermiideal}\, \ee to lowest order in $T/T_F^0$. As in
Sec.~\ref{ssec:atomenergy}, corrective terms due to the
interactions may be obtained in powers of $k_F^0 |a|$. The method
for doing so is outlined in App.~\ref{app:fermicorrect}. The
result is \be S = k_B \pi^2 N
\frac{T}{T_F^0}\left(1-\frac{64}{35\pi^2}k_F^0 |a|\right)
\label{eqn:fermientfinal}\ee to lowest order in $T/T_F^0$.  For
values of $k_F^0 |a| \lesssim 1/2$, as we have assumed, the
correction is less than 10\%.

\section{Positive scattering length: molecule-atom mix}
\label{sec:positive}

We next consider the case of positive scattering length, for which
there is a mixture of atoms and molecules, as in the experiment of
Ref.~\cite{greiner2003}.  In the following, it shall be assumed
that the molecular mean field acts on the atoms, but that the
effect of the atomic mean field on the molecules is negligible.
This is justified for small fractions of atoms, which is the case
experimentally.  The effect of the atomic and molecular mean
fields on \emph{themselves}, respectively, shall be calculated to
first order in the interactions, as was done in the previous
section for negative scattering length.

\subsection{Above the Molecular Condensation Temperature}
\subsubsection{Molecules}
\label{sssec:moleculesabove}

The entropy and kinetic, potential, and interaction energies of
the molecules may be calculated by similar methods to those
presented explicitly in Sec.~\ref{sec:negative}.  The local
density approximation for the energy in phase space is \be
E(r,p)=\frac{p^2}{2m_m}+\frac{1}{2}m_m\omega^2 r^2 +
g_{\mathrm{mm}}n_m(r) \label{eqn:phasespacemolabove}\ee One may
take the mean field density profile to be the non-interacting one,
{\it i.e.}, $n_m(r)=\frac{1}{2} n_a^{(0)}(r)$ (as in
Eq.~(\ref{eqn:density})) with all $a$ subscripts changed to $m$,
where the factor of one half is to account for the number of spin
states.  Then the resulting energies are:
\bea E_{\mathrm{kin}}&=& \frac{3}{2}N_m k_B T
\left[1-\frac{\zeta(3)}{16}\left(\frac{T^0_{\mathrm{BEC}}}{T}\right)^3 \right],
\label{eqn:KE4}\\
E_{\mathrm{pot}}&=& \frac{3}{2}N_m k_B T \left[1
+\frac{3\zeta(3)\gamma}{10\sqrt{2}}
\left(\frac{T^0_{\mathrm{BEC}}}{T}\right)^{5/2}\right.\nonumber\\
&&\left.
-\frac{\zeta(3)}{16}\left(\frac{T^0_{\mathrm{BEC}}}{T}\right)^3 \right],\label{eqn:POT4}\\
E_{\mathrm{int}}&=&N_m k_B T\frac{3\zeta(3)}{10\sqrt{2}}\,\gamma
\left(\frac{T^0_{\mathrm{BEC}}}{T}\right)^{5/2},\label{eqn:INT4}\eea
to cubic order in $T^0_{\mathrm{BEC}}/T$ and leading order in the
mean field in Eq.~(\ref{eqn:phasespacemolabove}), where \be \gamma
\equiv \left[\frac{N_m/N_a}{24\pi^3\zeta(3)}\right]^{1/6} k_F^0
|a|=\sqrt{\frac{k_B T^0_{\mathrm{BEC}}}{\pi E_B}}\,
,\label{eqn:gamma}\ee and \be E_B\equiv \frac{\hbar^2}{m_a a^2}\ee
is the approximate binding energy of the molecule.
Appendix~\ref{app:chempot} gives the perturbative expression for
the molecular fugacity, which is useful in obtaining the energies,
in that one must eliminate the chemical potential in order to
obtain energies depending on temperature and atomic/molecular
number alone.

Note that the large temperature non-interacting limit is given by
$E_{\mathrm{tot}}=E_{\mathrm{kin}}+E_{\mathrm{pot}}=3 N_m k_B T$.
This agrees with a simple calculation based on the density of
states: $E_{\mathrm{tot}}=\int dE E^3
e^{\beta(\mu_m-E)}/(\hbar\omega)^3=3 N_m k_B T +
\mathcal{O}(T^3)$. The expansions given by
Eqs.~(\ref{eqn:KE4}),~(\ref{eqn:POT4}), and~(\ref{eqn:INT4}) are
accurate to 0.66\%, 2.4\%, and 6.4\% for $k_F^0|a|=0.5$ and
$T/T_{\mathrm{BEC}}^0=1$. For higher temperature or weaker
interaction strength, they are accurate to better than 1\%.

The physically measurable quantity \be
\frac{E_{\mathrm{int}}}{E_{\mathrm{kin}}}=
\frac{\zeta(3)}{5\sqrt{2}}\left(\frac{T^0_{\mathrm{BEC}}}{T}\right)^{5/2}\gamma\,
,\ee to leading order, in contrast to the power law for Bose
condensed molecules, as shall be discussed in
Sec.~\ref{ssec:below}.

The entropy is calculable by similar techniques as were used to
obtain the energies, either from the combinatoric expression
similar to the calculation of Sec.~\ref{ssec:atomentropy}, or from
the grand potential (see, for example, Ref.~\cite{carr36}).

\subsubsection{Atoms}
\label{sssec:atomsabove}

%
\begin{figure}[t]
\begin{center}
\epsfxsize=8cm \leavevmode \epsfbox{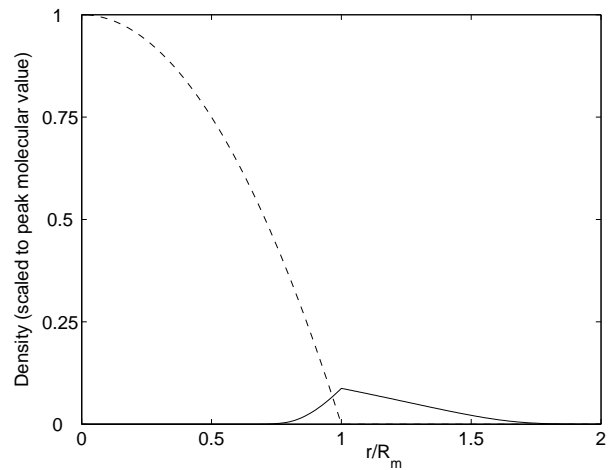}
\caption{\label{fig:2} Shown is the density profile of a Fermi gas
(solid curve) in the presence of a molecular condensate (dashed
curve), the latter being in the Thomas-Fermi limit with $R_m$ the
Thomas-Fermi radius. The atoms occupy a spherical shell outside
the molecular region.  Note that the parameters of the JILA
experiment~\cite{greiner2003} have been used for
$T/T^0_{\mathrm{BEC}}=0.25$ with the molecule--molecule and
atom-molecule scattering lengths derived in
Ref.~\cite{petrov2003b}.}
\end{center}
\end{figure}

The energy of the atoms in phase space is \be
E(r,p)=\frac{p^2}{2m_a}+\frac{1}{2}m_a\omega^2 r^2 + g
\frac{n_a(r)}{2}+ g_{\mathrm{am}}n_m(r)\, .
\label{eqn:phasespaceatomsabove}\ee Note the inclusion of the
molecular mean field.  The atom-atom and atom-molecule interaction
energies are \bea E^{\mathrm{aa}}_{\mathrm{int}}&=&g\int d^3r
\left(\frac{n_a(r)}{2}\right)^2
\, ,\\
E^{\mathrm{am}}_{\mathrm{int}}&=&g_{\mathrm{am}}\int d^3r\,
n_a(r)n_m(r) \, ,\eea respectively, while the kinetic and
potential energies are calculated in analogy with
Eqs.~(\ref{eqn:KE}) and~(\ref{eqn:PE}). One finds \bea
E_{\mathrm{kin}}&=&N_a k_B T
\left[\frac{3}{2}+\frac{3}{64}\frac{N_a\zeta(3)}{N_m}
\left(\frac{T^0_{\mathrm{BEC}}}{T}\right)^3\right]\, ,\label{eqn:KE5}\\
E_{\mathrm{pot}}&=&N_a k_B T \left[\frac{3}{2}+
\frac{N_a}{N_m}\frac{3\zeta(3)}{64}
\left(\frac{T^0_{\mathrm{BEC}}}{T}\right)^3\right.\nonumber\\
&&\left.+\gamma\,\zeta(3)\left(\frac{N_a}{N_m}\frac{3}{16}
+\frac{2\sqrt{3}}{5}\right)
\left(\frac{T^0_{\mathrm{BEC}}}{T}\right)^{5/2} \right]
,\:\:\:\:\:\label{eqn:POT5}\\
E^{\mathrm{aa}}_{\mathrm{int}}&=&N_a k_B T\frac{N_a}{N_m}
\frac{\zeta(3)}{8}\gamma\left(\frac{T^0_{\mathrm{BEC}}}{T}\right)^{5/2}\, ,
\label{eqn:INT5}\\
E^{\mathrm{am}}_{\mathrm{int}}&=&N_a k_B T
\frac{2\sqrt{3}\zeta(3)}{5}\gamma\left(\frac{T^0_{\mathrm{BEC}}}{T}\right)^{5/2}\,
,\label{eqn:INT6}\eea
where the expansions have been made to cubic order in the small
parameter $T^0_{\mathrm{BEC}}/T$ and to lowest order in the atomic
and molecular mean fields in Eq.~(\ref{eqn:phasespaceatomsabove}).
Appendix~\ref{app:chempot} gives the perturbative expressions for
the atomic and molecular fugacities, which are useful in obtaining
the energies.  Note that, as in Sec.~\ref{sssec:moleculesabove},
the large temperature non-interacting limit is given by
$E_{\mathrm{tot}}=E_{\mathrm{kin}}+E_{\mathrm{pot}}=3 N_a k_B T$.
The expansions given by
Eqs.~(\ref{eqn:KE5}),~(\ref{eqn:POT5}),~(\ref{eqn:INT5})
and~(\ref{eqn:INT6}) are accurate to 0.006\%, 1\%, 4\%, and 4\%
for $k_F^0|a|=0.5$ and $T/T_{\mathrm{BEC}}^0=2$.  For higher
temperature or weaker interaction strength, the accuracy rapidly
improves.  In the most extreme limits, e.g.
$T/T_{\mathrm{BEC}}^0=1$, the accuracies are on the order of 10\%
to 50\%.

The correct virial theorem must be modified from that of
Eq.~(\ref{eqn:virial1}): \be
2E_{\mathrm{kin}}-2E_{\mathrm{pot}}+3E^{\mathrm{aa}}_{\mathrm{int}}
+2E^{\mathrm{am}}_{\mathrm{int}}=0\, . \label{eqn:virial2}\ee Note
that, according to the above expressions, the condition for
chemical equilibrium, $2\mu_a=-E_B +\mu_m$, which we have
\emph{not} assumed here~\cite{kokkelmans2004}, would give \be
\frac{N_m}{N_a}=2\sqrt{\zeta(3)}\left(\frac{T^0_{\mathrm{BEC}}}{T}\right)^{3/2}
e^{\beta E_B /2}\, \ee for $T^0_{\mathrm{BEC}}/T\ll 1$.

The entropy of the atoms may be calculated by integrating
numerically over the general expression for the entropy density
given by Eq.~(\ref{eqn:genent}) with \be u\equiv
\beta[\mu_a-\frac{1}{2}m_a\omega^2r^2-\frac{1}{2}g\, n_a(r) -
g_{\mathrm{am}}n_m(r)]\, .\ee  An expansion may also be developed
in order to determine a perturbative expression in the Boltzmann
limit in the same fashion as was done for the energies.

\subsection{Below the Molecular Condensation Temperature}
\label{ssec:below}

\subsubsection{Molecules}
\label{sssec:moleculesbelow}

The chemical potential of the molecules may be calculated from the
Thomas-Fermi profile $n_m^{TF}(r)$ to be \be \mu_m =
\left(\frac{3\cdot
5}{\sqrt{2}\,2^4\,\pi}\right)^{2/5}(g_{\mathrm{mm}}N_m)^{2/5}(m_m\omega^2)^{3/5}\,
.\ee The energy in phase space is given by the energy of
Boguliubov quasiparticle excitations (see~\cite{carr36} and
references therein): \be E(r,p)\equiv
\left\{\sqrt{\displaystyle\frac{p^2}{2m_m}
\left[\frac{p^2}{2m_m}+2g_{\mathrm{mm}}n^{TF}_m(r)\right]}
\quad|r|\leq R_m \atop{\displaystyle\frac{p^2}{2m_m}+\frac{m_m
\omega^2r^2}{2} -\mu_m \quad\,\,\,\,\,\,\,\,\,\, |r|>R_m\,
.}\right .\, \ee The energies may then be solved for as \be
E=\frac{8
N_m}{\pi\zeta(3)}\left(\frac{T}{T^0_{\mathrm{BEC}}}\right)^3 k_B
T\,f(\beta \mu_m)\, ,\ee
where the integrals for the kinetic, potential, and interaction
energies are defined respectively by
\begin{widetext}
\bea f_{\mathrm{kin}}(\beta \mu_m) &\equiv& (\beta
\mu_m)^{3/2}\left\{ \int_0^1
dx\,x^{2}\int_0^{\infty}\frac{dy\,y^{3/2}} {\exp\sqrt{y[y+2 \beta
\mu_m(1-x^2)]}-1} + \frac{3\sqrt{\pi}}{4}\int_1^{\infty}dx\,x^{2k}
g_{5/2}[e^{\beta \mu_m(1-x^2)}]\right\}\, , \label{eqn:kin4}\\
f_{\mathrm{pot}}(\beta \mu_m) &\equiv& (\beta
\mu_m)^{5/2}\left\{\int_0^1 dx\,x^{4}
\int_0^{\infty}\frac{dy\,y^{1/2}}{\exp\sqrt{y[y+2 \beta
\mu_m(1-x^2)]}-1} +\sqrt{\frac{\pi}{2}}\int_1^{\infty}dx\,x^{4}
g_{3/2}[e^{\beta \mu_m(1-x^2)}]\right\}\, ,\label{eqn:pot4}\\
f_{\mathrm{int}}(\beta \mu_m) &\equiv& (\beta
\mu_m)^{5/2}\left\{\int_0^1 dx\,x^{2}(1-x^2)
\int_0^{\infty}\frac{dy\,y^{1/2}}{\exp\sqrt{y[y+2 \beta
\mu_m(1-x^2)]}-1}\right\} \, . \label{eqn:int4}\eea
\end{widetext}
Note that the virial
theorem, Eq.~(\ref{eqn:virial1}), holds.

The entropy has been calculated elsewhere in detail from the exact
density of states~\cite{carr36}: \bea
\rho(\epsilon)=[\mu_{m}^2/(\pi\hbar^3\omega^3)]
\{2\sqrt{2}z\tanh^{-1}[\sqrt{2z}/(1+z)
]\nonumber\\
+4z^{3/2}-\sqrt{2}z^2 [\pi+2\tan^{-1}
((1-z)/\sqrt{2z})]\nonumber\\
+(1+z)^2 [\theta_0 -\sin(4\theta_0)/4]\}\, ,\,\,\,\,\eea where
$z\equiv \epsilon/\mu_{m}$ is the rescaled energy and
$\theta_0\equiv\cos^{-1}(1/\sqrt{1+z})$. The result is \be S=k_B
\frac{N_{m}}{\zeta(3)}\left(\frac{T}{T^0_{\mathrm{BEC}}}\right)^3
G(\beta\mu_{\mathrm{m}})\, ,\label{eqn:sbose}\ee where \bea
G(\beta \mu_{\mathrm{m}})\equiv (\beta\mu_m)^3 \int_0^{\infty}dz
f(z)\left[\frac{\beta\mu_m z} { e^{\beta\mu_m z}-1}
\right.\nonumber\\
\left. - \ln\left(1-e^{-\beta\mu_m z}\right)\right] \,
,\label{eqn:G}\eea and \be f(z)\equiv
\frac{(\hbar\omega)^3}{\mu_{\mathrm{m}}^2}\rho(\epsilon)\ee so as
to make the units explicit.  Note that $G(1)=8.32$.

\subsubsection{Atoms}
\label{sssec:atomsbelow}

%
\begin{figure}[t]
\begin{center}
\epsfxsize=8cm \leavevmode \epsfbox{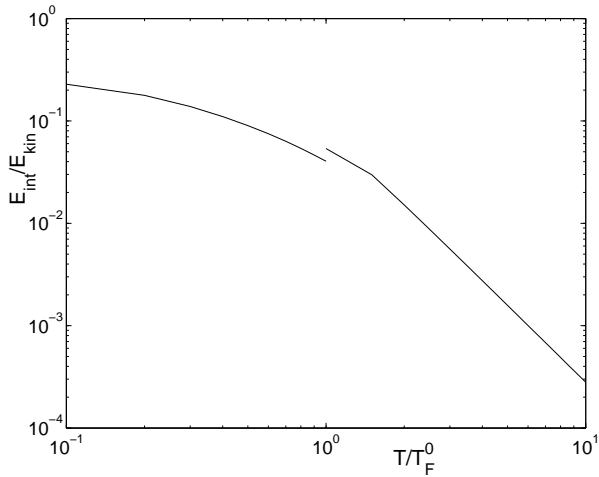}
\caption{\label{fig:3} {\it Atomic Fermi gas.}  Shown is the ratio
of the interaction energy to the kinetic energy as a function of
the degeneracy $T/T_F^0$, all in the presence of the molecular
mean field. Since the atoms are pushed to the outside of the
molecular mean field in space, the degeneracy cannot be extracted
in a simple way from the momentum distribution; this plot serves
as an alternative method to determine their degeneracy. The left
hand curve is calculated for a Bose-condensed molecular mean
field, {\it i.e.}, $T \ll T_{\mathrm{BEC}}^0$, while the right
hand one is calculated in the opposite limit. Note that $T_F^0=
1.53 (N_a/N_m)^{1/3} T_{\mathrm{BEC}}^0$; for the parameters shown
here, $T_F^0= 1.11\,T_{\mathrm{BEC}}^0$.}
\end{center}
\end{figure}

%
\begin{figure}[t]
\begin{center}
\epsfxsize=8cm \leavevmode \epsfbox{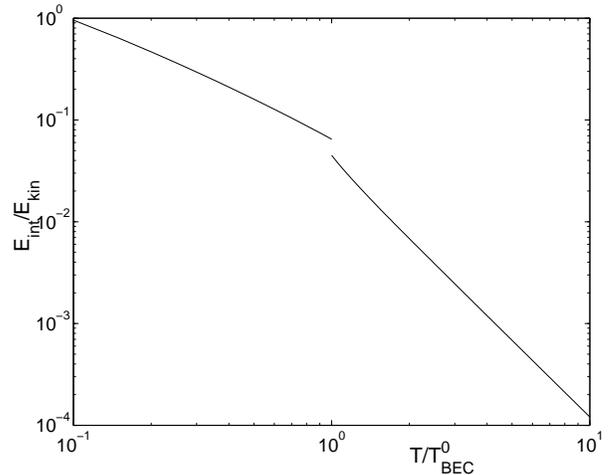}
\caption{\label{fig:4} {\it Molecular Bose gas.} Shown is the
ratio of the interaction energy to the kinetic energy as a
function of $T/T_{\mathrm{BEC}}^0$. This ratio may be used as a
signal of condensation since, for the molecules, the power law of
$E_{\mathrm{int}}/E_{\mathrm{kin}}\propto
(T/T^0_{\mathrm{BEC}})^{\delta}$ changes for $\delta \simeq -5/2$
above the phase transition to $\delta \simeq -1 $ below. The left
hand curve is calculated for a Bose-condensed molecular mean
field, {\it i.e.}, $T \ll T_{\mathrm{BEC}}^0$, while the right
hand one is calculated in the opposite limit. The parameters of
the JILA experiment have been used for plotting
purposes~\cite{greiner2003}.}
\end{center}
\end{figure}

The energy of the atoms in phase space is \be E(r,p) =
\frac{p^2}{2m_a}+\frac{1}{2}m_a\omega^2 r^2 +
g_{\mathrm{am}}n_m^{TF}(r) + \frac{1}{2}g n_a(r) \, .\ee Inserting
the Thomas-Fermi profile for the molecular density, one finds \be
E(r,p)\equiv \left\{\displaystyle\frac{p^2}{2m_a}-\frac{5 m_a
\omega^2 r^2}{2} -3\mu_m + \frac{g n_a(r)}{2} \quad|r|\leq R_m
\atop{\displaystyle\frac{p^2}{2m_a}+\frac{m_a \omega^2r^2}{2} +
\frac{g n_a(r)}{2} \quad\,\,\,\,\,\,\,\,\,\, |r|>R_m\, .}\right
.\, \ee Thus the atoms experience an \emph{expulsive} harmonic
potential~\cite{carr30} due to the molecular mean field and are
pushed to the outside of the molecular cloud.  The effect is
illustrated in Fig.~\ref{fig:2}.  To calculate the atomic density
profile to first order in the interactions, one follows similar
procedures as outlined in previous sections.  The result for
$n_a^{(0)}(r)$ is given by Eq.~(\ref{eqn:density}) with \be
u\equiv \left\{\beta\left(\mu_a+\frac{5}{2}m\omega^2 r^2
-3\mu_m\right) \quad|r|\leq R_m
\atop{\beta\left(\mu_a-\frac{1}{2}m\omega^2 r^2\right)
\quad\,\,\,\,\,\,\,\,\,\, |r|>R_m\, .}\right .\, \ee

Unlike in the other cases studied for atoms, we did not find a
simple way to expand the Bose functions in the integrands of the
energies. Therefore the energies are left determined up to an
integration which may be performed numerically for a given
parameter set, as in Eqs.~(\ref{eqn:kin4}) -~(\ref{eqn:int4}).

The entropy of the atoms may again be calculated by integrating
numerically over the general expression for the entropy density
given by Eq.~(\ref{eqn:genent}) with \be u\equiv
\beta[\mu_a-\frac{1}{2}m_a\omega^2r^2-\frac{1}{2}g\, n_a(r) -
g_{\mathrm{am}}n_m^{TF}(r)]\ee.

\section{Application to Experiments}
\label{sec:application}

The entropies and energies calculated in Secs.~\ref{sec:negative}
and~\ref{sec:positive} are useful in understanding experiments. In
the following, two examples are given. In
Figs.~\ref{fig:3}-\ref{fig:5} and Table~\ref{tab:1}, the
parameters of the recent JILA molecular BEC
experiment~\cite{greiner2003} are used throughout.

\subsection{Evidence of molecular condensation and atom-molecule temperature}

The dependence of the ratio of the interaction energy to the
kinetic energy on temperature is a useful experimental observable.
It is obtained via time-of-flight measurements with and without
interactions; the interactions can be rapidly switched on or off
via a Feshbach resonance~\cite{bourdel2003}.  This ratio allows
one to determine the degeneracy in the case of negative scattering
length (see Fig.~\ref{fig:1}). In the case of positive scattering
length, when one has an atom-molecule mixture, it can again be
used to calculate the temperature of the atoms. This is
particularly important as their temperature dependence cannot be
determined in a simple way from their momentum distribution, due
to the fact that the atoms occupy a spherical shell outside the
region of the molecular mean field (see Fig.~\ref{fig:2}). In
Fig.~\ref{fig:3} is shown this ratio, calculated for $T$ both
above and below $T_{\mathrm{BEC}}^0$, from the integral equations
of Secs.~\ref{sssec:atomsabove} and~\ref{sssec:atomsbelow}. In the
case of molecules, the dependence of the ratio of interaction to
kinetic energy on degeneracy signals the onset of condensation.
Above $T_{\mathrm{BEC}}^0$,
$E_{\mathrm{int}}/E_{\mathrm{kin}}\propto
(T/T^0_{\mathrm{BEC}})^{-5/2}$, while below
$E_{\mathrm{int}}/E_{\mathrm{kin}}\propto
(T/T^0_{\mathrm{BEC}})^{-1}$. The result is illustrated in
Fig.~\ref{fig:4}, as calculated from the energy equations of
Secs.~\ref{sssec:moleculesabove} and~\ref{sssec:moleculesbelow}.
Both figures use the parameters of the JILA
experiment~\cite{greiner2003}.  Note that the fact that the left
hand and right hand curves in Figs.~\ref{fig:3} and~\ref{fig:4} do
not meet is due to our use of the Thomas-Fermi approximation for
the mean field of the molecular condensate.  The use of this
approximation changes the form of the left hand curves as
$T\rightarrow T^0_{\mathrm{BEC}}$ from below; the right hand
curves are exact.

\subsection{A New Entropic Cooling Method}

The BCS transition temperature is given by~\cite{gorkov1} \be
T_{\mathrm{BCS}}/T_F^0 = 0.277
\exp\left(-\frac{\pi}{2k_F^0|a|}\right)\, .\label{eqn:bcsTc}\ee
For $k_F^0|a|=1/2$, the temperature necessary to achieve BCS is
$T_{\mathrm{BCS}}=0.012 \, T_F^0$. Present evaporative cooling
methods have been unable to achieve temperatures below about
$T=0.05 T_F$~\cite{greiner2003}. Recently, it was suggested that
adiabatic transition across a Feshbach resonance could be used to
transform a molecular BEC into a very cold Fermi gas, of
sufficiently temperatures to achieve BCS~\cite{carr36}. This is
due to the fact that the entropy of fermions is proportional to
$T$ while that of bosons is proportional to $T^3$. This scheme
requires evaporative cooling of the molecular BEC. The typical
maximal degeneracy achievable in an atomic BEC experiment is
$T/T_{\mathrm{BEC}}^0\simeq 1/4$.

%
\begin{figure}[t]
\begin{center}
\epsfxsize=8cm \leavevmode \epsfbox{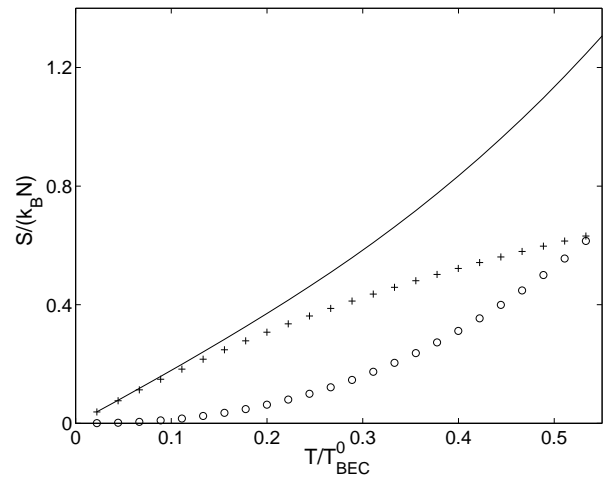}
\caption{\label{fig:5} Shown is the total entropy of an
atom-molecule mixture (solid curve), as well as the separate
contributions of the atoms (plus signs) and molecules (open
circles).  Clearly, for low temperature, the atoms carry the
majority of the entropy, which is what makes the entropic cooling
scheme outlined in Table~\ref{tab:1} effective. Note that the
parameters of the JILA experiment~\cite{greiner2003} have been
used.}
\end{center}
\end{figure}

Consideration of an atom-molecule mixture presents another
alternative.  Rather than trying to lower the temperature of the
system by evaporative cooling, one may instead attempt to decrease
the entropy.  Since entropy is held constant across the Feshbach
resonance, this amounts to decreasing the final temperature of the
Fermi gas.  In Fig.~\ref{fig:5} is shown the total entropy of an
atom-molecule mixture for the JILA parameters~\cite{greiner2003}
according to the expressions derived in the preceeding sections.
As illustrated in the figure, the atoms contribute the majority of
the total entropy when the Bose gas is degenerate. It has been
demonstrated experimentally that the atoms can be selectively
removed from an atom--molecule mixture~\cite{xu2003}.  We
therefore propose the following cooling scheme:
\begin{enumerate}
    \item A pseudo-spin-$1/2$ Fermi gas is evaporatively cooled to
degenerate temperatures for a small negative scattering length
($k_F^0|a|\ll 1$).
    \item A uniform external magnetic field is
    changed across a Feshbach resonance so as to cause the atoms to be
    transformed into molecules, slow enough to ensure thermal
    equilibrium. The scattering length is now small and positive
    ($k_F^0|a|\ll 1$, or, equivalently, $n^{1/3} a\ll 1$).
    \item A small number of atoms remain unpaired~\cite{greiner2003,regal2004}.
    These atoms are suddenly removed from the system, thereby reducing
    the entropy.
    \item The system is swept adiabatically back across the
    Feshbach resonance to a small, negative scattering length.  All
    molecules dissociate into atoms.  The resulting Fermi gas has a
    reduced temperature.
    \item The sequence is repeated until
    temperatures needed to obtain a BCS transition are achieved.
\end{enumerate}
In Table~\ref{tab:1} is presented a realization of this scheme,
utilizing parameters similar to those of Ref.~\cite{greiner2003},
where typically 85\% of the atoms are converted into molecules.
After three sweeps through the resonance, the final temperature of
$T = 0.0010 T_F^0$ is so low as to be out of the range of the
semi-classical approximation. Detailed requirements on the
adiabatic transition times and loss rates in the system have been
presented elsewhere~\cite{carr36}.

We note that, for the purposes of the table, it was assumed that
$g_{\mathrm{am}}=0.9 g$ and $g_{\mathrm{mm}}=0.3 g$, where
$g_{\mathrm{am}}$ and $g_{\mathrm{mm}}$ are the couplings for
atom--molecule and molecule--molecule
interactions~\cite{petrov2003b}.  This was necessary in order to
properly treat the effects of the atomic and molecular mean
fields.

\begin{table}[t]
\caption{\label{tab:1}An application of entropic cooling.
The left two columns depict the degeneracy and total number of fermionic
atoms for negative scattering length.  After adiabatic tuning via a Feshbach
resonance to positive scattering length (right facing arrow),
84 \% of the atoms are
transformed into bosonic molecules~\cite{greiner2003}.  In the two right
hand columns are shown the resulting entropy and degeneracy.  The remaining atoms are then suddenly
removed~\cite{xu2003}, thereby decreasing the entropy of the system, as shown in the third
column of the next row.  After tuning back
(left hand arrow), the temperature is reduced.  After several repetitions
of this process, the degeneracy is high enough to achieve a weakly interacting
BCS transition.
}
\begin{ruledtabular}
\begin{tabular}{lcccc }
$T/T_F$ & $N_a$ & Switching &Entropy $(k_B)$& $T/T_{\mathrm{BEC}}$ \\
\hline
0.100&  200,000 & $\rightarrow$  & 217,000 & 0.640\\
0.0504&  167,000 & $\leftarrow$  & 91,300  & 0.640\\

0.0504&  167,000 & $\rightarrow$ & 91,300 & 0.380\\
0.0147&  139,000 & $\leftarrow$  & 22,300 & 0.380\\
0.0147&  139,000 & $\rightarrow$ & 22,300 & 0.120\\
0.00103&  116,000 & $\leftarrow$ & 1,301  & 0.120 \\
\end{tabular}
\end{ruledtabular}
\end{table}

\section{Conclusions}
\label{sec:conclusion}

The entropy and kinetic, interaction, and potential energy of
fermionic atoms and bosonic molecules were calculated for
temperatures both above and below $T_F^0$ and
$T_{\mathrm{BEC}}^0$.  The effect of the atomic and molecular mean
fields were taken into account both perturbatively in analytic
expansions and self-consistently, as illustrated in the figures.
Below the condensation temperature, the molecular mean field was
assumed to be in the Thomas-Fermi limit and the thermal cloud was
neglected. The general method of performing these thermodynamic
calculations was outlined.

Two applications of the calculations were then proposed.  Firstly,
it was shown that the ratio of the kinetic to interaction energy
is sufficient to determine the degeneracy of the atoms.  The same
ratio, for the molecules, may be used to indicate the onset of
Bose condensation, as the power law dependence on the degeneracy
changes from an exponent of $-5/2$ to $-1$.  Secondly, it was
suggested that an adaption of entropic cooling~\cite{carr36} could
be used to cool a Fermi gas to sufficiently high degeneracy so as
to achieve a weakly interacting BCS transition, without the need
for evaporative cooling of either the fermionic atoms or the
bosonic molecules.  This scheme involves tuning adiabatically back
and forth through a Feshbach resonance: on the positive scattering
length side, the sudden removal of remaining unpaired atoms
corresponds to a large reduction in entropy, which, after return
to the negative scattering length, results in a decreased
temperature.  For example, for a conversion efficiency of atoms to
molecules of 85\%, as in Ref.~\cite{greiner2003}, and starting
with a Fermi gas of degeneracy $T/T_F^0=0.1$, after three sweeps
the degeneracy is increased by two orders of magnitude to
$T/T_F^0=0.001$ with a loss of less than half of the
atoms~\cite{loss}.

We note that entropic cooling was used successfully for the first time
in a recent experiment~\cite{bartenstein2004}.

We thank Deborah Jin, Marcus Greiner, Cindy Regal, Gora
Shlyapnikov, and Yvan Castin, for useful discussions.  We are
grateful to Jochen Wachter, Jinx Cooper, and Simon Gardiner for a
careful reading of the manuscript.  The support of the U.S.
Department of Energy, Office of Basic Energy Sciences via the
Chemical Sciences, Geosciences and Biosciences Division is
acknowledged. L.~D.~Carr also acknowledges the support of the
National Science Foundation via grant no.~MPS-DRF 0104447

\appendix
\section{Correction to the
kinetic, potential, and interaction energies for a weakly interacting Fermi gas}
\label{app:fermicorrect}

Beginning with the general expressions for the energies given by
Eqs.~(\ref{eqn:KE}) -~(\ref{eqn:INT}), one may integrate over
momentum to obtain \bea E_{\mathrm{kin}}&=&A_{\mathrm{kin}}
\int_0^{\infty} dy\, y^{1/2} g_{5/2}\left(- e^{u}\right)
\, , ~\label{eqn:KE2}\\
E_{\mathrm{pot}}&=&A_{\mathrm{pot}}\int_0^{\infty} dy\, y^{3/2}
g_{3/2}\left(- e^{u}\right)
\, , ~\label{eqn:PE2}\\
E_{\mathrm{int}}&=&A_{\mathrm{int}}\int_0^{\infty} dy\, y^{1/2}
\left[\frac{1}{2}g_{3/2}\left(- e^{u}\right)\right]^2 \,
,~\label{eqn:INT2}\eea where \be u\equiv \beta\left[\mu_a -y
-\frac{1}{2}g n_a(r)\right]\, ,\ee \be y\equiv \frac{1}{2}\beta
m_a \omega^2 r^2 \, ,\ee and \bea
A_{\mathrm{kin}}&=&-\frac{18 N}{\sqrt{\pi}}\left(\frac{T}{T_F^0}\right)^3 k_B T\, ,\\
A_{\mathrm{pot}}&=&-\frac{12 N}{\sqrt{\pi}}\left(\frac{T}{T_F^0}\right)^3 k_B T\, ,\\
A_{\mathrm{int}}&=&-\frac{6
N}{\pi}\left(\frac{T}{T_F^0}\right)^{7/2} k_B T\, k_F^0 |a|\, .
\eea

To resolve the integrals to first order in the interactions,
it is sufficient to assume a mean field given by the
non-interacting density profile of Eq.~(\ref{eqn:density}).  Then, one may expand
the Bose functions in the integrands of Eqs.~(\ref{eqn:KE2}) -~(\ref{eqn:INT2})
as \bea g_n\left\{-\exp\left[\beta\mu_a-y-\alpha
g_{3/2}\left(-e^{\beta\mu_a-y}\right)\right]\right\}\nonumber\\
\sim  g_n[-\exp(\beta\mu_a-y)]\nonumber\\
-\alpha g_{3/2}[-\exp(\beta\mu_a-y)] g_{n-1}[-\exp(\beta\mu_a-y)]
\, , \label{eqn:gexpand}\eea where \be \alpha \equiv
|a|\sqrt{\frac{2m}{\hbar\pi\beta}}\, .\label{eqn:alpha}\ee is the
small expansion parameter.  The maximum value of the Bose function
in the non-interacting density occurs for $y=0$ for positive
$\beta\mu_a$.  In the limit $\beta\mu_a \gg 1$, the Bose function
may be expanded as~\cite{landau3} \bea g_n\left(-
e^{\upsilon}\right)=-\frac{\upsilon^n}{\Gamma(n)}
\left[\frac{1}{n}+\frac{\pi^2}{6}(n-1)\frac{1}{\upsilon^2}
\right.\nonumber\\ \left.
+\frac{7\pi^4}{360}(n-1)(n-2)(n-3)\frac{1}{\upsilon^4}\right]\,
.\label{eqn:boseexpand} \eea Thus, in the degenerate regime, \be
\max\left[g_{3/2}\left(-e^{\beta\mu_a-y}\right)\right]\sim
\left(\frac{1}{\beta\mu_a}\right)^{3/2}\, . \ee  One may therefore
take as a condition for the expansion of Eq.~(\ref{eqn:gexpand}):
\be \beta\mu_a \gg \frac{1}{(\beta E_B)^{1/3}}\, ,\ee where
$E_B\equiv \hbar^2/(m_aa^2)$ is the approximate binding energy of
the molecule.

Using the integral relations
\bea g_4(z)=\frac{2}{\sqrt{\pi}}\int_0^{\infty}
dy\, y^{1/2}g_{5/2}(z e^{-y})\, ,\nonumber\\
g_3(z)=\frac{2}{\sqrt{\pi}}\int_0^{\infty} dy\, y^{1/2}g_{3/2}(z
e^{-y})\, ,\label{eqn:myint}\eea {\it etc.},  one may resolve the
first term in the expansion of Eq.~(\ref{eqn:gexpand}) exactly. An
additional useful relation, necessary for calculating the
potential energy by integration by parts, is \be \int dy
\,g_{n}\left(q e^{-y}\right)=-g_{n+1}\left(q e^{-y}\right)\, . \ee

To resolve the order $\alpha$ term of Eq.~(\ref{eqn:gexpand}), one
notes that there are two regimes of integration: (a) $y\ll \beta
\mu_a$ and (b) $y \gg \beta \mu_a$.  For a degenerate Fermi gas,
so that $\beta \mu_a \gg 1$, the two leading contributions to the
integrals are obtained in regime (a).  Therefore, in order to find
the result to order $(T/T_F^0)^2$, for which the two leading order
terms are required, it suffices to perform the integral by the use
of the expansion given in Eq.~(\ref{eqn:boseexpand}).  Expanding
around small $y$ and integrating, one obtains a final result
similar to Eqs.~(\ref{eqn:KE3}) -~(\ref{eqn:INT3}) but dependent
on the chemical potential $\mu_a$.

To obtain a self-consistent correction, the relation between the
chemical potential and the Fermi temperature can be calculated
from the equation for the total number of fermions \be N = n_s
\int \frac{d^3r d^3p}{(2\pi\hbar)^3} \nu(r,p) \,  ,
~\label{eqn:atomNum}\ee where $n_s=2$ is the number of spin
states.  One finds, by the same methods as outlined above, that
\bea \frac{\mu}{k_B T_F^0} &=& \left(1
- \frac{512}{315\pi^2}k_F^0 |a|\right)\nonumber\\
&&-\left(\frac{\pi^2}{3}+\frac{16}{315}k_F^0 |a|\right)\left(\frac{T}{T_F^0}\right)^2\,
\label{eqn:atomNum2}\eea
to second order in $T/T_F^0$ and first order in $k_F^0 |a|$.
Eliminating the chemical potential from the equations for the energies via substitution
of Eq.~(\ref{eqn:atomNum2}), one obtains Eqs.~(\ref{eqn:KE3}) -~(\ref{eqn:INT3}).

\section{Correction to the entropy for a weakly interacting Fermi gas}
\label{app:fermicorrect2}

Equation~(\ref{eqn:entropy1}) may be conveniently rearranged as
\be S=2 k_B \sum_{\{r,p\}}[ \nu x
+\ln(1+e^{-x})]\label{eqn:entropy2}\ee where \be x\equiv
\beta(E(r,p)-\mu)\, .\ee  Assuming the semiclassical local density
approximation, one can make the change of variables \bea
u&\equiv&\beta\left[\mu_a-V(r)-U_{mf}(r)\right]
\, ,\label{eqn:udef}\\
z&\equiv& \beta \frac{p^2}{2m_a}\, ,\label{eqn:zdef}\\
S&\equiv& \int d^3 r\, S(r)\, .\eea
Then, integrating the second term in Eq.~(\ref{eqn:entropy2}) by parts, one finds
\bea S(r)&=&A \int_0^{\infty}dz \,z^{1/2}\frac{(z-u)+\frac{2}{3}z}{1+\exp(z-u)}\, ,\\
 A&\equiv& \frac{k_B}{2\pi^2} \left(\frac{2m_a}{\beta\hbar^2}\right)^{3/2}\, .\eea
This integral may be resolved as a sum of two Bose
functions~\cite{huang1} of form $g_n(x)$, as given in
Eq.~(\ref{eqn:genent}).

\section{Dependence of the Fugacity on Temperature in the Boltzmann limit}
\label{app:chempot}

In order to calculate the thermodynamic quantities for the
atom-molecule mixture at $T>T_{\mathrm{BEC}}^0$ in
Secs.~\ref{sssec:moleculesabove} and~\ref{sssec:atomsabove}, the
following perturbative expansions of the fugacities are useful.
Starting with Eq.~(\ref{eqn:atomNum}), and including the mean
fields, the relations are

\bea e^{\beta\mu_m} &\simeq&
\zeta(3)\left(\frac{T_{\mathrm{BEC}}^0}{T}\right)^3
+\frac{4\sqrt{2}\,\gamma}{15}\left(\frac{T_{\mathrm{BEC}}^0}{T}\right)^{11/2}\nonumber\\
&&-\frac{[\zeta(3)]^2}{8}
\left(\frac{T_{\mathrm{BEC}}^0}{T}\right)^6,\,\,\,\,\,\eea for the
molecular fugacity and
\bea e^{\beta\mu_a} &\simeq& \frac{\zeta(3)}{2}\frac{N_a}{N_m}
\left(\frac{T_{\mathrm{BEC}}^0}{T}\right)^3
\nonumber\\
&&+\gamma\frac{N_a}{N_m}\left(\frac{\sqrt{3}}{5}
-\frac{1}{8}\frac{N_a}{N_m}\right)
\left(\frac{T_{\mathrm{BEC}}^0}{T}\right)^{11/2}\nonumber\\
&&+\frac{[\zeta(3)]^2}{32}\frac{N_a^2}{N_m^2}
\left(\frac{T_{\mathrm{BEC}}^0}{T}\right)^6\eea for the atomic
fugacity, where $\gamma\propto k_F^0|a|$ is defined in
Eq.~(\ref{eqn:gamma}).


\begin{thebibliography}{10}

\bibitem{randeria1995}
M. Randeria, {\em Bose-Einstein Condensation} (Cambridge
University Press,
  U.K., 1995), Chap.~15, pp.\ 355--392.

\bibitem{holland2001}
M. Holland, S.~J. J. M.~F. Kokkelmans, M.~L. Chiofalo, and R.
Walser, Phys.
  Rev. Lett. {\bf 87},  120406  (2001).

\bibitem{timmermans2001}
E. Timmermans, K. Furuya, P.~W. Milonni, and A.~K. Kerman, Phys.
Lett. A {\bf
  285},  228  (2001).

\bibitem{ohashi2002}
Y. Ohashi and A. Griffin, Phys. Rev. Lett. {\bf 89},  130402
(2002).

\bibitem{density}
It is also possible to obtain such superfluid states by increasing
the density
  and/or changing the interaction strength. Here, as we consider a fixed total
  number of atoms in a harmonic trap with weak interactions, it is the
  temperature which plays the most important role in obtaining BEC or BCS.

\bibitem{greiner2003}
M. Greiner, C.~A. Regal, and D.~S. Jin, Nature {\bf 426},  437
(2003).

\bibitem{zwierlein2003}
M.~W. Zwierlein, C.~A. Stan, C.~H. Schunck, S.~M.~F. Raupach, S.
Gupta, Z.
  Hadzibabic, and W. Ketterle, Phys. Rev. Lett. {\bf 91},  250401  (2003).

\bibitem{jochim2003b}
S. Jochim, M. Bartenstein, A. Altmeyer, S. Riedl, C. Chin, J.~H.
Denschlag, and
  R. Grimm, Science {\bf 302},  2102  (2003).

\bibitem{regal2004}
C.~A. Regal, M. Greiner, and D.~S. Jin, Phys. Rev. Lett. {\bf 92},
040403
  (2004).

\bibitem{kinast2004}
J. Kinast, S.~L. Hemmer, M.~E. Gehm, A.~Turlapov, and J.~E.
Thomas, Phys. Rev.
  Lett. {\bf 92},  150402  (2004).

\bibitem{bourdel2004}
T. Bourdel, L. Khaykovich, J. Cubizolles, J. Zhang, F. Chevy, M.
Teichmann,
  S.~J. J. M.~F. Tarruell, L.~Kokkelmans, and C. Salomon, 2004, e-print
  cond-mat/0403091.

\bibitem{vogels1}
J.~M. Vogels, C.~C. Tsai, R.~S. Freeland, S.~J. J. M.~F.
Kokkelmans, B.~J.
  Verhaar, and D.~J. Heinzen, Phys. Rev. A {\bf 56},  R1067  (1997).

\bibitem{holland1}
M.~J. Holland, B. DeMarco, and D.~S. Jin, Phys. Rev. A {\bf 61},
53610
  (2000).

\bibitem{koehler2003b}
T. Kohler, T. Gasenzer, and K. Burnett, Phys. Rev. A {\bf 67},
013601
  (2003).

\bibitem{ohashi2003}
Y. Ohashi and A. Griffin, Phys. Rev. A {\bf 67},  033603  (2003).

\bibitem{petrov2003b}
D.~S. Petrov, C. Salomon, and G.~V. Shlyapnikov, 2003, e-print
  cond-mat/0309010.

\bibitem{kokkelmans2004}
S.~J. J. M.~F. Kokkelmans, G.~V. Shlyapnikov, and C. Salomon,
Phys. Rev. A {\bf
  69},  031602  (2004).

\bibitem{holland2004}
M.~J. Holland, C. Menotti, and L. Viverit, 2004, e-print
cond-mat/0404234.

\bibitem{diener2004}
R.~B. Diener and T.-L. Ho, 2004, e-print cond-mat/0404517.

\bibitem{falco2004}
G.~M. Falco and H.~T.~C. Stoof, Phys. Rev. Lett. {\bf 92},  130401
(2004).

\bibitem{drummond2004}
P.~D. Drummond and K.~V. Kheruntsyan, Coherent molecular
bound-states of bosons
  and fermions near a Feshbach resonance, 2004, e-print cond-mat/0404429.

\bibitem{andreev2004}
A.~V. Andreev, V. Gurarie, and L. Radzihovsky, 2004,
  e-print cond-mat/0404724.

\bibitem{bourdel2003}
T. Bourdel, J. Cubizolles, L. Khaykovich, K.~M.~F. Magalhaes,
S.~J. J. M.~F. Kokkelmans, G.~V.
  Shlyapnikov, and C. Salomon, Phys. Rev. Lett. {\bf 91},  020402  (2003).

\bibitem{ohara1}
K.~M. O'Hara, S.~L. Hemmer, M.~E. Gehm, S.~R. Granade, and J.~E.
Thomas,
  Science Express {\bf 298},  2179  (2002).

\bibitem{ho2004}
T.-L. Ho, Phys. Rev. Lett. {\bf 92},  090402  (2004).

\bibitem{carr36}
L.~D. Carr, G.~V. Shlyapnikov, and Y. Castin, Phys. Rev. Lett.
{\bf 92},
  150404  (2004).

\bibitem{huang1}
K. Huang, {\em Statistical Mechanics}, 2nd ed. (John Wiley \&
Sons, New York,
  NY, 1987).

\bibitem{narrow}
In experiments on $^6$Li, a negligible fraction of atoms were
observed after crossing the resonance from negative to positive
scattering length, in contrast to the case of $^{40}$K.  This has
been conjectured to be due to the differing widths of the Feshbach
resonances.  See~\cite{depalo2004} for further discussion on the
matter.

\bibitem{xu2003}
K. Xu, T. Mukaiyama, J.~R. Abo-Shaeer, J.~K. Chin, D.~E. Miller,
and W.
  Ketterle, Phys. Rev. Lett. {\bf 91},  210402  (2003).

\bibitem{chevy2002}
F. Chevy, V. Bretin, P. Rosenbusch, K.~W. Madison, and J.
Dalibard, Phys. Rev.
  Lett. {\bf 88},  250402  (1997).

\bibitem{hadzibabic2003}
Z. Hadzibabic, S. Gupta, C.~A. Stan, C.~H. Schunck, M.~W.
Zwierlein, K. Dieckmann,
  and W. Ketterle, Phys. Rev. Lett. {\bf 91},  160401  (2003).

\bibitem{bartenstein2004}
M. Bartenstein, A. Altmeyer, S. Riedl, S. Jochim, C. Chin, J.~H.
Denschlag, and
  R. Grimm, Phys. Rev. Lett. {\bf 92},  120401  (2004).

\bibitem{depalo2004}
S. De~Palo, M.~L. Chiofalo, M.~J. Holland, and S.~J. J. M.~F.
Kokkelmans, 2004,
  e-print cond-mat/0404672.

\bibitem{anisotropic}
For most thermodynamic calculations, a generalization to the
anisotropic case
  can be made simply by substituting $\bar{\omega}\equiv
  (\omega_x\omega_y\omega_z)^{1/3}$ for $\omega$ in all expressions.

\bibitem{combescot2001}
R. Combescot, Europhys. Lett. {\bf 55},  150  (2001).

\bibitem{dalfovo1}
F. Dalfovo, S. Giorgini, L.~P. Pitaevskii, and S. Stringari, Rev.
Mod. Phys.
  {\bf 71},  463  (1999).

\bibitem{metikas}
See~\cite{metikas2004} and references therein for theoretical
treatments which go beyond the mean field approximation.

\bibitem{landau3}
L.~D. Landau and E.~M. Lifshitz, {\em Statistical Physics, Part I}
(Reed
  Educational and Professional Publishing Ltd., Boston, Massachussets, 1980),
  Vol.~5.

\bibitem{carr30}
L.~D. Carr and Y. Castin, Phys. Rev. A {\bf 66},  063602  (2002).

\bibitem{gorkov1}
L.~P. Gor'kov and T.~K. Melik-Barkhudarov, Sov. Phys. JETP {\bf
13},  1018
  (1961).

\bibitem{loss}
This estimate excludes other loss processes. Note that the
semiclassical
  approximation begins to break down for temperatures $T\ll 0.01 T_F$, so that
  the final temperature of $T=0.001 T_F$ must be taken as qualitative. However,
  even temperatures of $T\lesssim 0.01 T_F$ are sufficient to achieve a weakly
  interacting BCS transition.

\bibitem{metikas2004}
G. Metikas, O. Zobay, and G. Alber, Phys. Rev. A {\bf 69},  043614
(2004).

\end{thebibliography}

\end{document}